\documentclass[aps,prb,preprint,groupedaddress,floatfix]{revtex4}
\usepackage{graphicx,amssymb}
\begin{document}
\title{Unraveling liquid polymorphism in silicon driven out-of-equilibrium}
\author{Caroline Desgranges and Jerome Delhommelle\footnote{Author to whom correspondence should be addressed: jerome.delhommelle@und.edu}}
\affiliation{Department of Chemistry, New York University, New York, New York 10003, United States}
\affiliation{Department of Chemistry, University of North Dakota, Grand Forks ND 58202, United States}

\date{\today}

\begin{abstract}
Using nonequilibrium molecular dynamics (NEMD) simulations, we study the properties of supercooled liquids of $Si$ under shear at $T=1060$~K over a range of densities encompassing the low-density liquid (LDL) and high-density liquid (HDL) forms. This enables us to generate nonequilibrium steady-states of the LDL and HDL polymorphs, that remain stabilized in their liquid forms for as long as the shear is applied. This is unlike the LDL and HDL forms at rest, which are metastable under those conditions and, when at rest, rapidly undergo a transition towards the crystal, {\it i.e.} the thermodynamically stable equilibrium phase. In particular, through a detailed analysis of the structural and energetic features of the liquids under shear, we identify the range of densities, as well as the range of shear rates, that give rise to the two forms. We also show how the competition between shear and tetrahedral order impacts the two-body entropy in steady-states of Si under shear. These results open the door to new ways of utilizing shear to stabilize forms that are metastable at rest and can exhibit unique properties, since, for instance, experiments on Si have shown that HDL is metallic, with no band gap, while LDL is semimetallic, with a pseudogap.
\end{abstract}

\maketitle

\section{Introduction}

Recent experimental and computational studies have revealed that the liquid state can also exhibit the phenomenon of polymorphism~\cite{poole1997polymorphic,harrington1997liquid,mishima1998decompression,koga2000first,franzese2001generic,kurita2004critical,roberts1996liquid}. Similarly to the solid state that has been known, for a long time, to exhibit different crystal structures or polymorphs, the liquid state can also exhibit different liquid polymorphs, with distinct densities, structures and entropies. In particular, different liquid forms have been shown to exist in atomic systems, including phosphorus~\cite{katayama2000first}, carbon~\cite{togaya1997pressure,glosli1999liquid}, silicon~\cite{sastry2003liquid,zhang2014liquid,jakse2007liquid,beye2010liquid}, in molecular fluids like water~\cite{mishima2000liquid,soper2000structures,poole1992phase,mishima1998relationship,abascal2010widom,liu2012liquid,palmer2014metastable,singh2016two,palmer2013liquid,palmer2018advances,limmer2011putative,limmer2013putative,palmer2018comment}, triphenyl phosphite~\cite{kurita2005abundance,tanaka2004liquid,tanaka2000general}, alcohols~\cite{huvs2014existence,desgranges2018communication} and in aqueous organic solutions~\cite{murata2013general}. In the case of silicon, the presence of two liquids below the melting point of Si, and the existence of a first order transition between the two, have been established by computer simulations~\cite{sastry2003liquid,beaucage2005liquid,beaucage2005nucleation} using a classical force field, known as the Stillinger-Weber (SW) potential~\cite{stillinger1985computer}. These two liquid forms, referred to as the low-density liquid (LDL) form and the high-density liquid (LDL) form, differ in density, structure and transport properties like diffusivity and viscosity. For instance, the density is found to be greater by $5$\% in HDL than in LDL, the coordination number also decreases from $4.9$ (HDL) to $4.24$ (LDL), and LDL is a network liquid with a high amount of tetrahedral order~\cite{sastry2003liquid}. Diffusivity is roughly two orders of magnitude smaller in LDL than in HDL~\cite{sastry2003liquid}. Similarly, LDL is much more viscous than HDL. Equilibrium molecular dynamics simulations~\cite{mei2013dynamic} have reported Green-Kubo calculations for the viscosity that showed, over the $1000-1100$~K temperature interval, dramatically greater viscosities for LDL. An important characteristic of LDL, and of the LDL-HDL transition, is that these have only been observed below the melting point, where the Si crystal is the thermodynamically stable phase and the two liquid forms LDL and HDL are metastable~\cite{debenedetti1996metastable}. LDL is indeed often thought as a precursor for the formation of amorphous Si at high supercooling and to crystal nucleation at low supercooling~\cite{beaucage2005liquid,beaucage2005nucleation,desgranges2011role}. This metastability was leveraged to observe in the transient regime the two liquid forms LDL and HDL in a pioneering experiment that subjected a crystal of Si to ultrashort optical pulses of femtosecond duration~\cite{beye2010liquid}. This triggered the melting of the crystal in LDL, that rearranged then into HDL. These structural changes were also accompanied by changes in the electronic properties, since LDL is semimetallic, with a pseudogap, while HDL is metallic, with no band gap. These observations pave the way for a control of liquid polymorphism via the use of an external field. Moreover, while the experimental observations were made on transient states and thus, on a very short timescale, a suitable choice of external perturbations could allow for the observation of liquid polymorphs in the steady-state, {\it i.e.} for as long as the external field is switched on. In particular, shear is often a very useful tool to probe nonequilibrium phase transitions including the solid-liquid transition~\cite{butler2002factors,delhommelle2004simulations,ramsay2016shear}, the liquid-liquid transition in a model system for methanol~\cite{desgranges2018communication} and, more generally, the nonequilibrium response of glasses~\cite{williams2010rheology} and supercooled liquids~\cite{abraham2012origin,desgranges2008rheology}.

The aim of this study is to address the following questions: (i) can nonequilibrium steady-states of the LDL and HDL forms be obtained by subjecting supercooled systems of Si to shear?, and (ii) how does the competition between shear and tetrahedral ordering impact the structure, rheology and entropy of Si under shear? For this purpose, we use nonequilibrium molecular dynamics (NEMD) simulation to study the response of supercooled Si over a wide range of shear rates and densities. The model used for Si is the classical SW model. While this force field does not allow for the calculation of electronic properties, the SW model provides a solid basis to analyze the nonequilibrium response of Si under shear since the LDL and HDL forms have been extensively characterized at rest and the model provides an accurate picture for the decrease in tetrahedral ordering with increasing density~\cite{sastry2003liquid,beaucage2005liquid,beaucage2005nucleation}. Building on our previous work using the SLLOD algorithm~\cite{evans2008non,delhommelle2004simulations,desgranges2008shear,desgranges2008rheology,desgranges2018communication}, we obtain liquid systems of Si that are driven out-of-equilibrium by the applied shear and remain in a steady-state for as long as shear is applied. Through a series of analyses of the variation of the structural, energetic and rheological properties of silicon under shear, we elucidate the conditions (density and applied shear rate) for which LDL and HDL are obtained in driven Si. We also unravel the interplay between tetrahedral order and the shear-induced structural changes that take place in Silicon under shear and characterize the nonequilibrium two-body entropy in steady-states of LDL and HDL under shear.

The paper is organized as follow. In the next section, we present the simulation method, model, structural and energetic analyses to characterize the properties of supercooled systems of $Si$ under shear. We then present the simulation results obtained at $T=1060$~K over a range of densities extending from $2.28$~g/cm$^3$ to $2.52$~g/cm$^3$ and compare the properties of supercooled liquids of Si under shear to the equilibrium data for the LDL and HDL polymorphs. In particular, we identify that the features characterizing the LDL and HDL at rest are found in steady-state liquids of Si under shear, provided that the applied shear rate remains sufficiently low. We also discuss how the competition between shear and tetrahedral order impacts the two-body entropy in steady-states of Si under shear, before drawing the main conclusions from this work in the last section.

\section{Simulation methods}
We use nonequilibrium molecular dynamics (NEMD)~\cite{morriss1989phase,evans2000note,todd2007homogeneous,ewen2018advances} to study the response of liquid silicon undergoing shear flow. In this work, we carry out simulations using an in-house code in the isothermal ensemble (NVT) with a number of silicon atoms set to $N=512$, a temperature of $T=1060~K$, for which prior work has shown that there is a liquid-liquid transition~\cite{sastry2003liquid,beaucage2005liquid}, and for $7$ values of the volume $V$ corresponding to densities ranging from $2.28~$g/cm$^3$ to $2.52$~g/cm$^3$, with a $0.04$~g/cm$^3$ interval. To model the interactions between Si atoms, we use the well-established Stillinger-Weber (SW) potential~\cite{stillinger1985computer}. As shown in previous work~\cite{sastry2003liquid,beaucage2005liquid,beaucage2005nucleation}, this model yields equilibrium pressures that are consistent with the results from {\it ab initio} calculations~\cite{zhao2016phase}. It is defined as the sum of a two-body term and of a three-body term. The pairwise term $u_2$ is given by
\begin{equation}
\begin{array}{llll}
u_2(r_{ij}) & = & A \epsilon(B(r_{ij}/\sigma))^{-p}-(r_{ij}/\sigma)^{-q}) \exp \left[ ((r_{ij}/\sigma-a)^{-1}\right] & (r_{ij}/\sigma)<a\\
u_2(r_{ij}) &= & 0 & (r_{ij}/\sigma)\ge a\\
\end{array}
\end{equation}
with the parameters $\epsilon=50$~kcal/mol, $\sigma=2.0951$~\AA, $A=7.04955627$, $B=0.602224558$, $p=4$, $q=0$ and $a=1.8$.
and of a three-body term $u_3$ written as
\begin{equation}
u_3(\mathbf{r_i},\mathbf{r_j},\mathbf{r_k})=\epsilon [h(r_{ij},r_{ik},\theta_{jik})+h(r_{ji},r_{jk},\theta_{ijk})+h(r_{ki},r_{kj},\theta_{ikj})]
\end{equation}
where the $h$ function is defined for $r<a$ as, {\it e.g.}, in the case of $h(r_{ij},r_{ik},\theta_{jik})$
\begin{equation}
h(r_{ij},r_{ik},\theta_{jik})= \lambda \exp [\nu(r_{ij}/\sigma-a)^{-1}+\nu(r_{ik}/\sigma-a)^{-1}] \times (cos \theta_{jik}+1/3)^2
\end{equation}
where $\theta_{jik}$ denotes the angle between vectors $\mathbf{r}_{ij}$ and $\mathbf{r}_{ik}$, subtended by vertex $i$, and where $\lambda=21$ and $\nu=1.2$.

We simulate a planar Couette flow in the $\mathbf{x}$ direction with a velocity gradient along the $\mathbf{y}$ direction using the SLLOD algorithm, together with the Lees-Edwards boundary conditions\cite{evans2008non}. The equations of motion for a $N$-particle system subject to a steady external shear rate $\gamma$ are given by
\begin{equation}
\begin{array}{lll}
\mathbf{\dot q_i} & = & {\mathbf{p_i} \over m}  + \gamma y_i \mathbf{e}_ x \\
\mathbf{\dot p_i} & = & \mathbf{F_i} -  \gamma p_{yi} \mathbf{e}_ x- \alpha \mathbf{p_i}\\
\end{array}
\label{EOM}
\end{equation}
In these equations, heat is dissipated via the use of a Gaussian thermostat~\cite{sarman1994extremum,dettmann1996hamiltonian}, which is the last term of the second line in Eq.~\ref{EOM}. $\alpha$ is the thermostat multiplier and is defined as
\begin{equation}
\alpha = -{  {\sum_{i=1}^N \mathbf{p_i . F_i} - \gamma {\sum_{i=1}^N p_{x,i} . p_{y,i} }} \over {\sum_{i=1}^N \mathbf{p_i}^2} }
\end{equation} 
Here, the rate at which work done on the system by the external field $\gamma$ is compensated by the rate at which heat is removed from the system by the thermostat. This allows the system to reach a steady state~\cite{evans2008non}. The choice of a Gaussian thermostat is not expected to impact the results, as the shear rates considered here are less than $1$ (in reduced units). Previous work on atomic fluids has shown that profile-unbiased thermostats~\cite{travis1995thermostats,bagchi1996profile}, such as configurational thermostats~\cite{delhommelle2001comparison,lue2002configurational,delhommelle2002correspondence,braga2005configurational,samoletov2010notes}, provide a more physical basis for heat dissipation in liquids subjected to shear rates greater than $1$ and allow for the onset of secondary flow profiles in strongly sheared liquid~\cite{delhommelle2003reexamination}.

We integrate the equations of motion with a five-value Gear predictor-corrector algorithm and a time step of  $1 \times 10^{-15}$~s. For each value of the shear rate $\gamma$, we start by running a first run of $2 \times 10^6$ time steps, or, in other words, a trajectory of $2$~ns, and check that the system has reached a steady state. Then, we perform an additional production run of $2 \times 10^6$ time steps to compute time averages of various physical properties of the system, including the two-body energy $u_2$, the three-body energy $u_3$, the viscosity $\eta$ and the pressure $P$. In particular, the viscosity $\eta$ can be calculated as~\cite{zhang1999kinetic}
\begin{equation}
\eta= -{{ \left< {P_{xy}} \right>} \over {\gamma}}
\label{etaNEMD}
\end{equation}
This method is very well suited to calculate transport properties in the steady-state, {\it i.e.} when a steady linear flow profile has developed across the fluid~\cite{evans2008non}. Other methods, such as the transient-time correlation function (TTCF) formalism~\cite{evans1988transient,desgranges2008molecular,desgranges2008rheology,hartkamp2012transient,desgranges2008shear}, apply when the response in the transient regime needs to be determined. Throughout this work, we use a system of reduced units for simulation parameters, such as the reduced shear rate $\gamma^*$, in which the unit length is set to $\sigma$, the energy unit to $\epsilon$ and the unit mass to $m$, the atomic mass of Si.

Moreover, we use different order parameters to study the system. First, the global order parameter $Q_6$, introduced by Steinhardt {\it et al.}~\cite{Steinhardt}, that measures the amount of crystalline order in a system
\begin{equation}
\label{invarglobal}
Q_6=\left[{{4\pi \over 13} \sum_{m=-6}^6 \left| {\sum_i \sum_j Y_{6m}(\hat{\mathbf r}_{ij}) \over \sum_i N_b(i)} \right| ^2}\right]^{1/2}
\end{equation}
where $\hat{\mathbf r}_{ij}$ is the unit vector joining two neighboring atoms $i$ and $j$, that are less than a distance of $4.1$~\AA~from each other (corresponding to the first minimum of the pair distribution function for the liquid), $Y_{6m}(\hat{\mathbf r}_{ij})$ is a spherical harmonics and $N_b(i)$ the number of neighbors for molecule $i$. $Q_6$ takes values greater than $0.3$ for crystal phases, and vanishes in the liquid~\cite{tenWolde}.
We also use the order parameter $q_t$ that quantifies the average amount of tetrahedral order~\cite{chau1998new,errington2001relationship}. It is calculated as an average of the local $q_t(i)$ over all atoms $i$ in the system with
\begin{equation}
q_t(i)=1- {3 \over 8} \sum_{j=1}^3 \sum_{k=j+1}^4 \left( {\cos \psi_{jk} + {1 \over 3}} \right)^2
\end{equation}
where $j$ and $k$ are two atoms among the $4$ nearest neighbors of atom $i$, and $\psi_{jk}$ is the angle formed by the line joining $i$ and $j$ and the line joining $i$ and $k$. Perfect tetrahedral order around an atom $i$ corresponds to $q_t(i)=1$.

\section{Results and discussion}

\begin{figure}
\begin{center}
\includegraphics*[width=7.5cm]{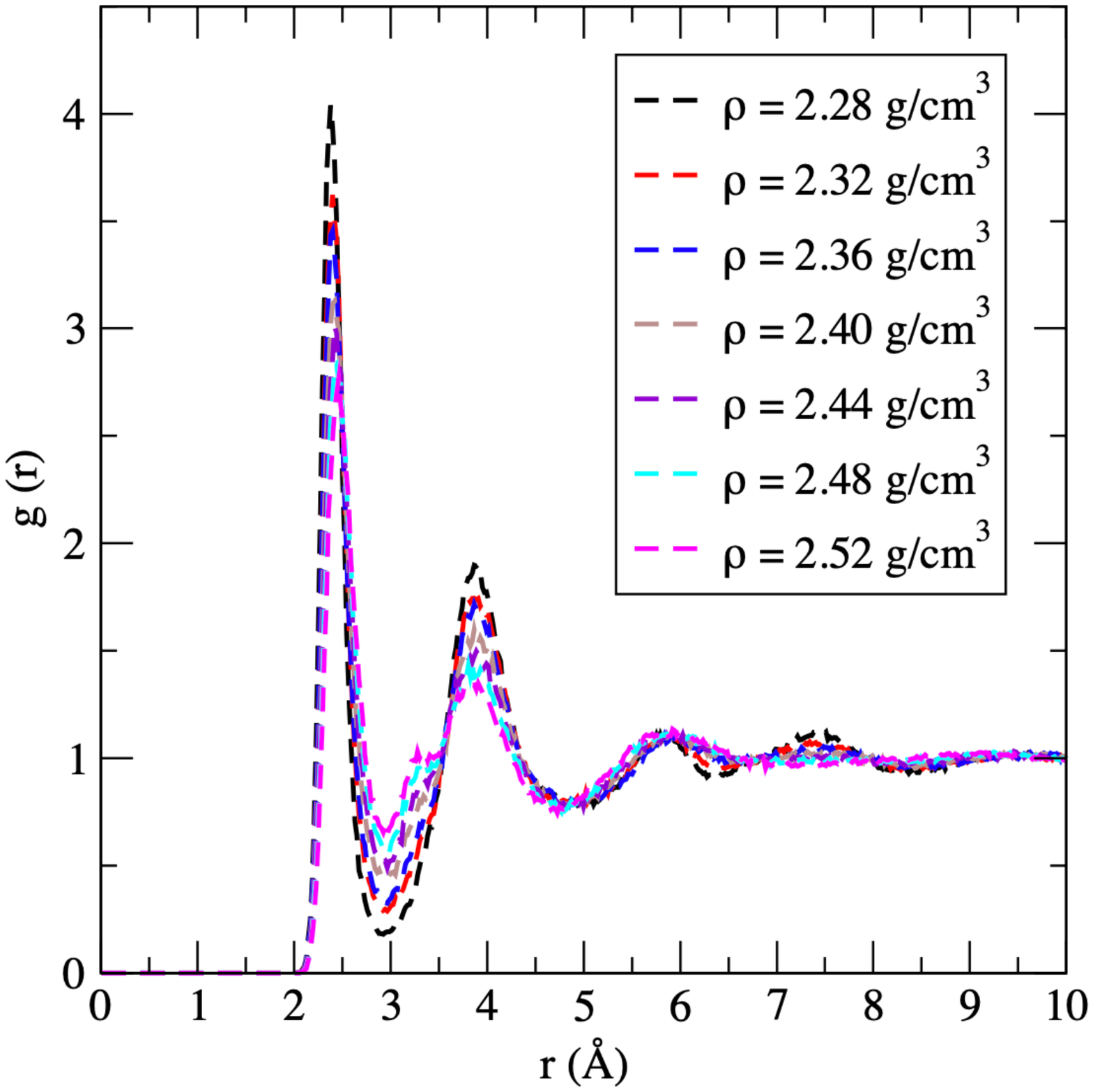}(a)
\includegraphics*[width=7.5cm]{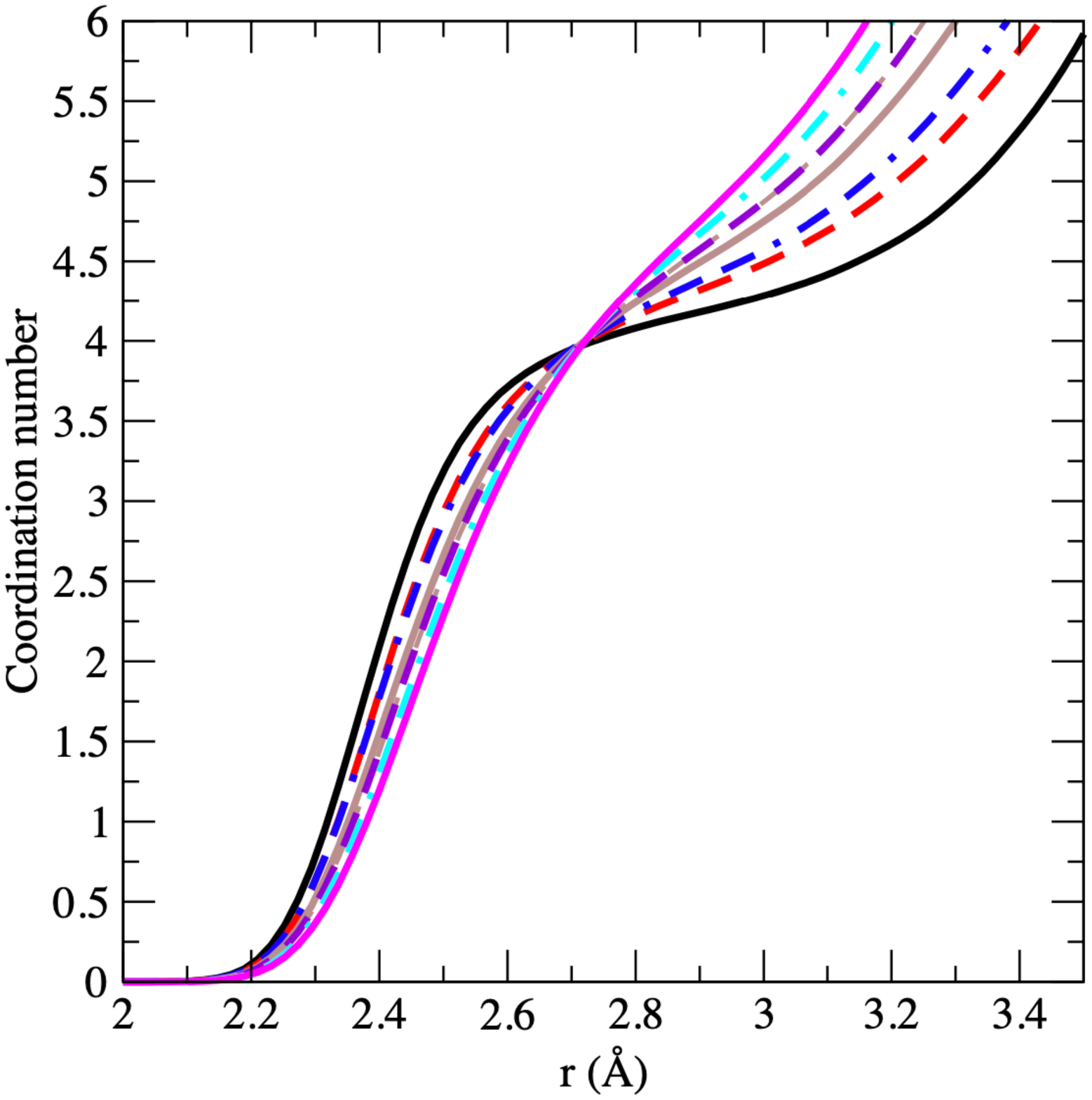}(b)
\end{center}
\caption{(a) Pair correlation function g(r) for densities ranging from $\rho=2.28~g/cm^3$ to $\rho=2.52~g/cm^3$. (b) Coordination number obtained by integrating g(r). Same caption as in (a).}
\label{Fig1}
\end{figure}

We start by commenting on the results obtained at rest ($\gamma=0$) as a function of density, since they provide a baseline to understand how shear impacts the properties of supercooled Si. For a temperature $T=1060$~K, previous work~\cite{sastry2003liquid,beaucage2005liquid} has shown that there is a first-order liquid-liquid transition from a high-density liquid (HDL) to a low-density liquid (LDL). Since this temperature is below the melting point, we emphasize that these states are metastable. This means that, at rest, they can, and eventually will, evolve with time. Here we show results at rest obtained after $4\times 10^6$ time steps or, equivalently, after $4$~ns.  We present in Fig.~\ref{Fig1}(a) the radial distribution function $g(r)$ for supercooled liquids of Si with densities ranging from $2.28$~g/cm$^3$ to $2.52$~g/cm$^3$. Examination of the evolution of the $g(r)$ as a function of density shows two distinct behaviors below and above a density of $2.4$~g/cm$^3$. The structural differences can best be seen by looking at the features of the first two peaks in the $g(r)$. For the lower densities, the first maximum is reached at about $2.38$~\AA~with a value close to $4$ for this distance. The first minimum is located at $2.95$~\AA, which delimits the first coordination shell to be between $2$~\AA~and $2.95$~\AA. The second peak is reached for $3.9$~\AA~with a maximum of about 2 (half of that in the first shell) and a second minimum located at $4.9$~\AA~, which means that the second coordination shell is between $2.95$~\AA~ and $4.9$~\AA. For densities greater than $2.4$~g/cm$^3$, the first maximum is reached for a distance of $2.42$~\AA~and a maximum probability of about $3$ ({\it i.e.} 33\% less than for the lower densities, and the first minimum can be seen for $2.97$~\AA). These features are in overall agreement with the structures found for LDL and HDL using {\it ab initio} molecular dynamics~\cite{zhao2015nature}, with, most notably, the increased height of the first peak of $g(r)$ in LDL and a narrowing of the distance between the first and second maximum for HDL when compared to LDL. Unlike for the lower densities, a shoulder develops around $3.4$~\AA~between the first minimum ($2.97$~\AA) and the second maximum ($3.89$~\AA). In fact, this shoulder starts to form for $\rho=2.40$~g/cm$^3$ and becomes more predominant as the density of the system increases up to $\rho=2.52$~g/cm$^3$. It is the signature of the onset of structural order, that has been referred to as medium-range order (MRO)~\cite{hui2003shoulder,tanaka2012bond}. We add that, for all systems, $g(r)$ converges towards one, showing that there is no long-range order. We also compute the value taken for $Q_6$ and find that, for all systems, $Q_6$ remains close to $0.014$ throughout the time interval spanned during the simulations, confirming that we have obtained metastable liquids over the entire range of densities.

To analyze further the structure of these liquids, we calculate the coordination number $N_c$ by integrating g(r) as $\int 4 \pi r^2 \rho g(r) dr$ and show the results in Fig.~\ref{Fig1}(b).  For the lower densities, we find $N_c=4.23$ in the first coordination shell, for $r<2.95$~\AA~(numbers given here for $\rho=2.28$~g/cm$^3$). This coordination number, together with the $g(r)$ reported in Fig.~\ref{Fig1}(a), indicate that the lower end of the density range corresponds to the low-density liquid (LDL) of Si. Indeed, previous studies~\cite{sastry2003liquid,beaucage2005liquid,beaucage2005nucleation} have shown that the LDL is characterized by a largely tetrahedral structure with a coordination number between $4.20$ and $4.24$ at $1050$~K in the $NPH$ ensemble.  On the other hand, for densities above $2.4$~g/cm$^3$, the coordination number is $N_c=4.83$ in the first coordination shell, {\it i.e.} $r<2.98$~\AA~(here on the example of $\rho=2.44$~g/cm$^3$). This is consistent with the larger $N_c$ reported for the high-density liquid (HDL)~\cite{sastry2003liquid,beaucage2005liquid,beaucage2005nucleation}  for $Si$. This establishes that the range of densities studied here covers the two liquid forms LDL to HDL found at rest in supercooled $Si$.

\begin{figure}
\begin{center}
\includegraphics*[width=7.6cm]{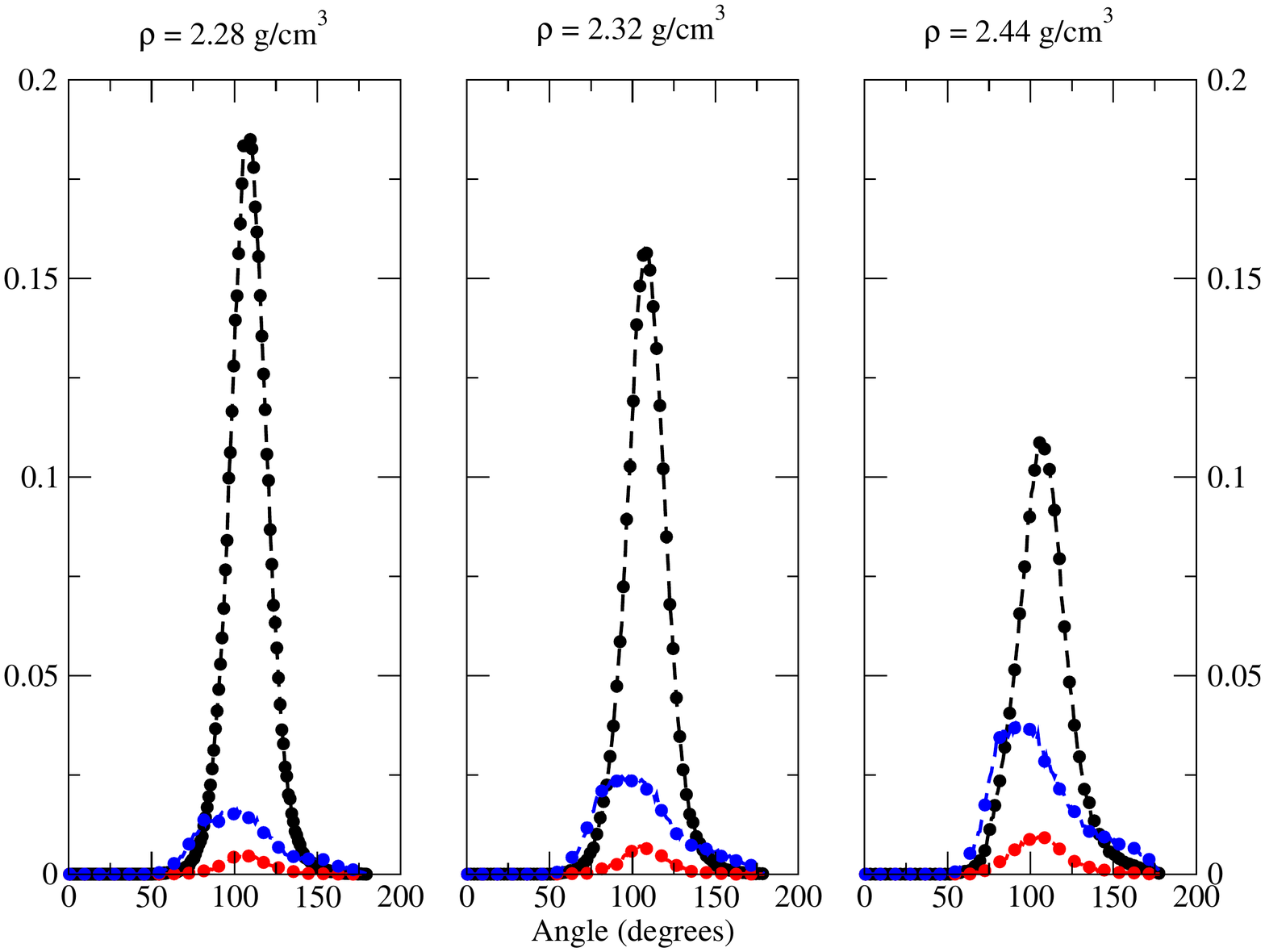}(a)
\includegraphics*[width=7.6cm]{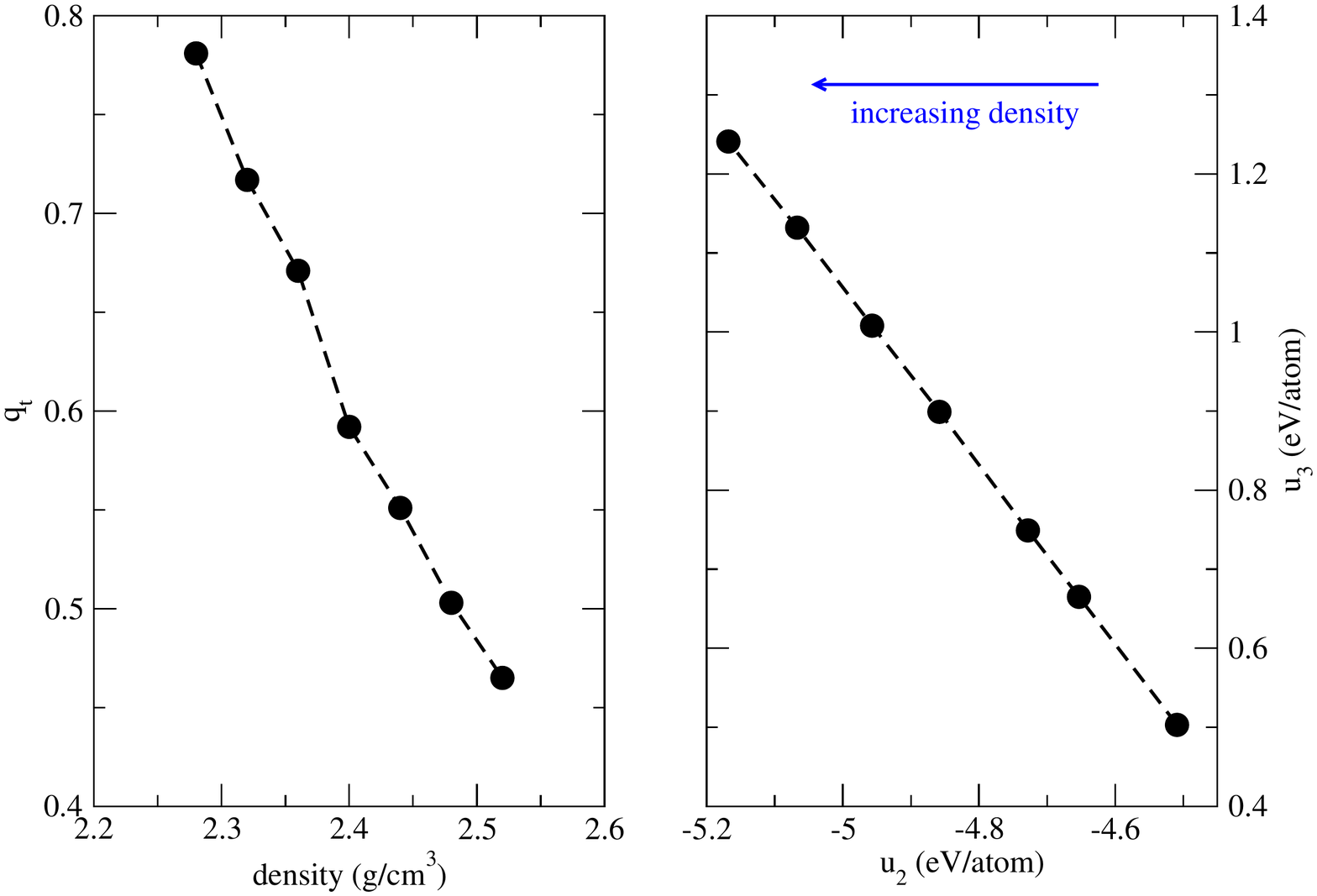}(b)
\includegraphics*[width=7.6cm]{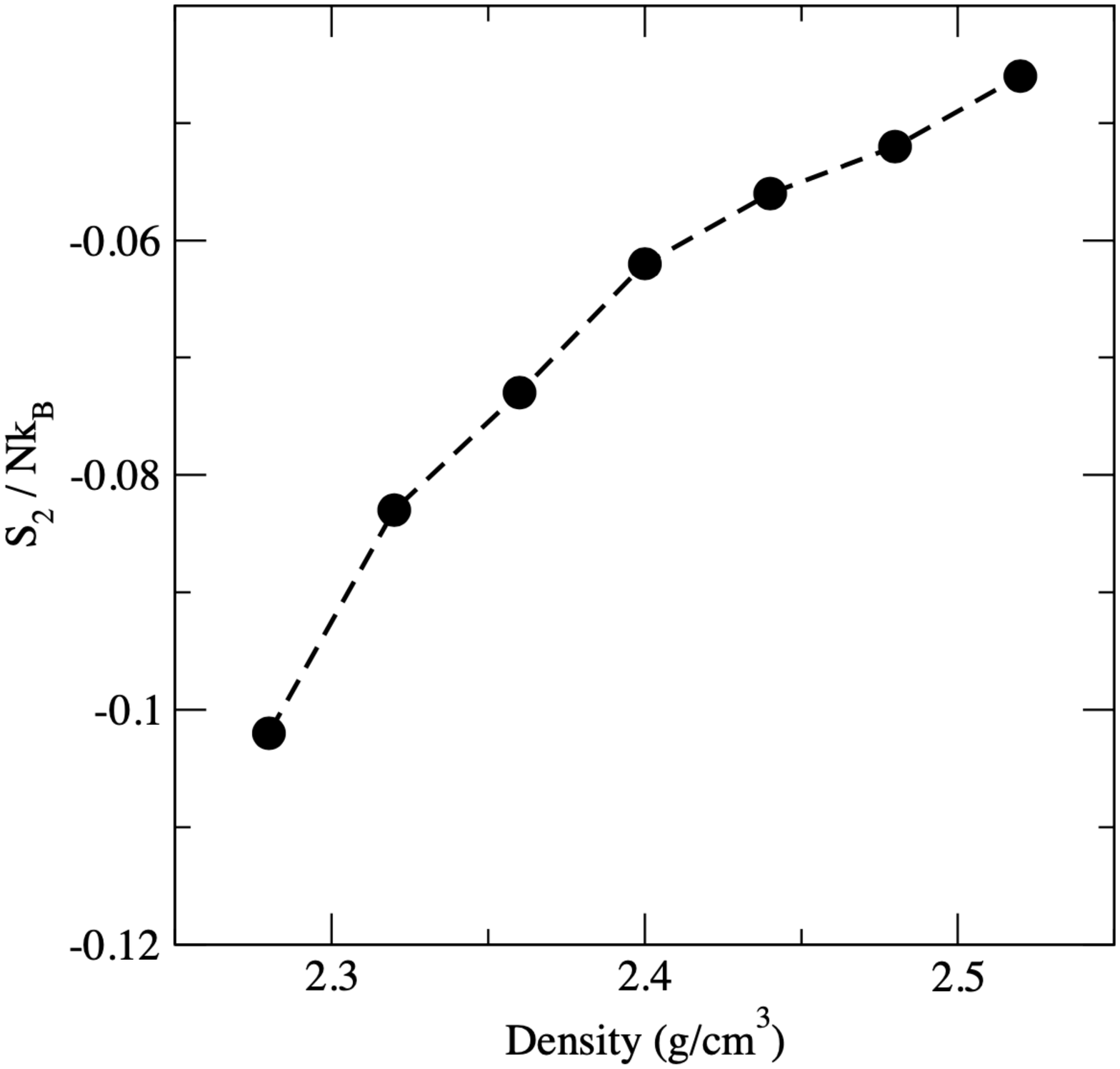}(c)
\end{center}
\caption{(a) Angle distribution for $\rho=2.28~g/cm^3$, $\rho=2.32~g/cm^3$ and $\rho=2.44~g/cm^3$. Distributions are shown for atoms with $3$ (in red), $4$ (in black) and $5$ (in blue) first neighbors. (b) Left: tetrahedral local order parameter $q_t$ as a function of density. Right: Variation of 2-body SW energy vs. 3-body SW energy for increasing density (the first filled circle in the right bottom corner is for $\rho=2.28~g/cm^3$). (c) Entropy $S_2$ as a function of density in supercooled liquids of Si at rest.}
\label{Fig2}
\end{figure}

We also consider other structural features and order parameters at rest to characterize in the next paragraphs how shear impacts the structure and properties of supercooled liquids of Si. In particular, we examine, as a function of density, the angle distributions for atoms with $3$, $4$ or $5$ neighbors within a sphere of a $2.75$~\AA~radius ({\it i.e.} the cutoff radius for the SW 3-body potential) and show the resulting plots in Fig.~\ref{Fig2}(a) for $\rho=2.28$~g/cm$^3$, $\rho=2.32$~g/cm$^3$ and $\rho=2.44$~g/cm$^3$. Fig.~\ref{Fig2}(a) shows that, for all densities, the most frequent number of first neighbors is $4$ (in black) with a distribution centered around the expected angle for a tetrahedral environment ($109.5^\circ$). However, the corresponding probability steadily decreases with density from $85$\% in the LDL ($\rho=2.28$~g/cm$^3$) to $64$\% in the HDL ($\rho=2.44$~g/cm$^3$). Instead, atoms with both $3$ and $5$ first neighbors become more frequent, from $5$\% and $9\%$ at $\rho=2.28$~g/cm$^3$ to $12$\% and $21$\% at $\rho=2.44$~g/cm$^3$, respectively. This finding is consistent with results from prior {\it ab initio} equilibrium molecular dynamics calculations on LDL and HDL~\cite{zhao2015nature}. The decrease in tetrahedral order with density can also be measured by the order parameter $q_t$, as seen in the left panel of Fig.~\ref{Fig2}(b). The loss of tetrahedral order results in a decrease of $q_t$ from about $0.78$ at low density ($\rho=2.28$~g/cm$^3$) to $0.46$  ($\rho=2.52$~g/cm$^3$). This is accompanied by a combined increase in the 3-body energy and decrease in 2-body energy. Indeed, the 3-body energy in the SW potential is purely repulsive and reaches a minimum of $0$ when there is a perfect tetrahedral environment around an atom. On the other hand, since a density increase results in a loss of tetrahedral order, the increase in 3-body energy with density is expected. Similarly, the decrease in the purely attractive 2-body energy is consistent with the increase in $N_c$ that results from the increase in density. Very interestingly, when at rest, both LDL and HDL liquids see their 2-body and 3-body energy fall onto the line shown in the right panel of Fig.~\ref{Fig2}(b). As we will see, this behavior differs markedly from what is observed under shear. Finally, we examine how entropy can be quantified in these highly nonequilibrium systems. Indeed, the evaluation of entropy out-of-equilibrium systems has recently drawn considerable interest for metastable liquids undergoing a nucleation process~\cite{desgranges2016free,desgranges2016free2,desgranges2017free,piaggi2017enhancing,desgranges2018crystal}, for systems driven out-of-equilibrium~\cite{zu2020information} and in active matter~\cite{martiniani2019quantifying}. Here we examine how the onset of tetrahedral ordering in LDL can be monitored by the decrease in the pair-correlation entropy $S_2$, defined as
\begin{equation}
\label{S2l}
S_2= -{\rho \over 2} \int_0^\infty \left[ g(r) \ln g(r) - \left(g(r)-1\right) \right] dr
\end{equation}
Fig.~\ref{Fig2}(c) shows that the enhanced structural features exhibited by the LDL form (see the $g(r)$ shown in Fig.~\ref{Fig1}(a)) results in a lower $S_2$ entropy than at high densities, {\it i.e.} for the HDL form. In other words, the results show that $S_2$ is sensitive enough, in the case of Si, to characterize the differences between LDL and HDL at rest and to serve as a baseline for the results under shear that we will discuss in the next paragraphs.

\begin{figure}
\begin{center}
\includegraphics*[width=16cm]{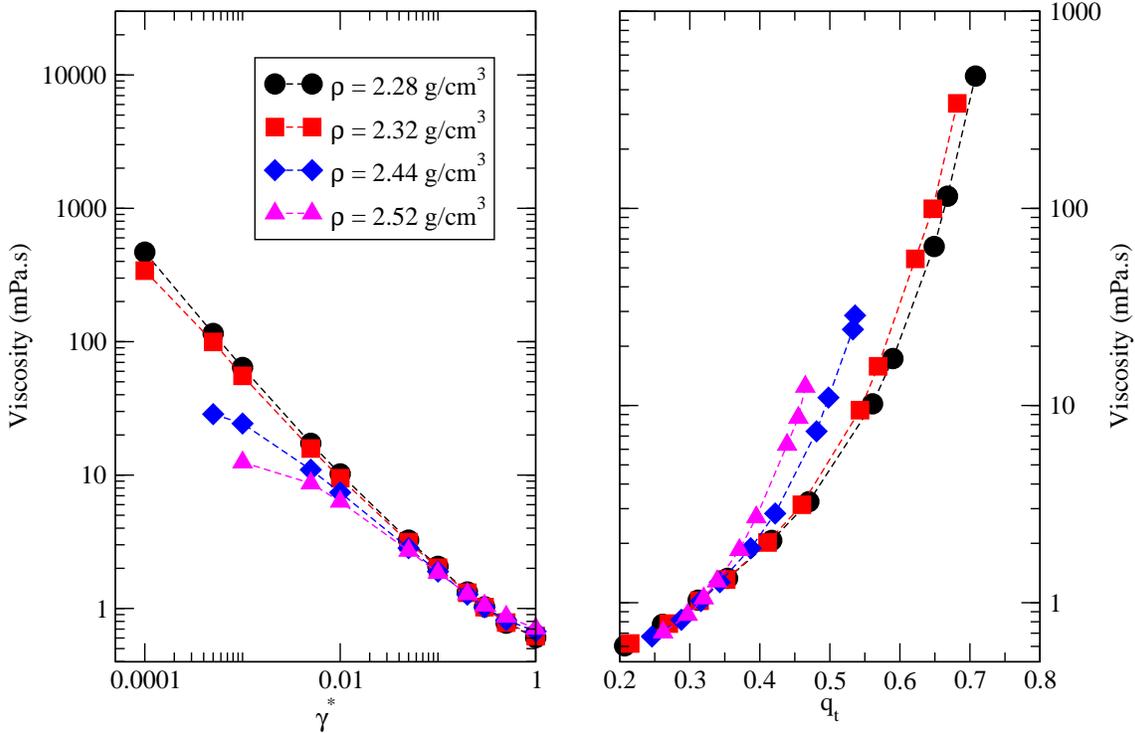}
\end{center}
\caption{(Left) Shear viscosity ($\eta$) as a function of shear rate ($\gamma^*$) for $\rho=2.28$~g/cm$^3$, $\rho=2.32$~g/cm$^3$, $\rho=2.44$~g/cm$^3$ and $\rho=2.52$~g/cm$^3$. (Right) Shear viscosity  against the tetrahedral local order parameter $q_t$ (lines are plotted as a guide to the eye).}
\label{Fig3}
\end{figure}

We now turn to the response of supercooled liquids of $Si$ when subjected to shear. Results shown in Fig.~\ref{Fig3} are obtained in the steady-state when a linear flow profile, with a slope equal to the imposed shear rate $\gamma^*$, has developed across the system. These systems remain liquid in the steady-state as a result of the constant input of mechanical energy exerted by the imposition of this constant shear rate (structural features of the liquids under shear are presented in Figs. 3-5). The shear viscosity of supercooled liquids of $Si$ are shown on the left panel of Fig.~\ref{Fig3} for $\rho=2.28$~g/cm$^3$, $\rho=2.32$~g/cm$^3$, $\rho=2.44$~g/cm$^3$ and $\rho=2.52$~g/cm$^3$. The liquids exhibit the expected shear-thinning behavior, with a decrease in shear viscosity with an increase in shear rate. The viscosity plots for the four densities become very similar for reduced shear rates beyond $0.01$. As shown on the right panel of Fig.~\ref{Fig3}, these also amount to very similar values of the order parameter $q_t$ for the three densities, which implies that shear rates beyond $0.01$ essentially wipe away any density-dependent structural features that can be seen in LDL and HDL systems at rest. On the other hand, when the reduced shear rate becomes lower than $1 \times 10^{-3}$, shear viscosities and $q_t$ both start to depend strongly on density and become much greater for $\rho=2.28$~g/cm$^3$ and $\rho=2.32$~g/cm$^3$ than for $\rho=2.44$~g/cm$^3$. For instance, for $\rho=2.44$~g/cm$^3$, the viscosity starts to reach the Newtonian plateau for shear rates of the order of $1 \times 10^{-3}$ with $\eta=28$~mPa.s. However, for the lower densities, the shear viscosity continues to increase as the shear rate decreases and reaches $460$~mPa.s ($\rho=2.28$~g/cm$^3$) and $\eta=340$~mPa.s ($\rho=2.32$~g/cm$^3$) for $\gamma^*=1 \times 10^{-4}$. To determine the zero-shear (Newtonian) viscosity, we fit the data for the shear-rate dependent viscosity $\eta_N$ using an Eyring model~\cite{jadhao2017probing}. We find that, in accord with prior work~\cite{mei2013dynamic}, the zero-shear rate viscosity for HDL is of a the order of a few tens mPa.S with {\it e.g.} $\eta_N=15$~mPa.s at $2.52$~g/cm$^3$, while the zero viscosity for LDL is two orders of magnitude greater with {\it e.g.} $\eta_N=1614$~mPa.s at $2.32$~g/cm$^3$.

Most notably, we find that $q_t$ is in excess of $0.7$  for $\gamma^*=1 \times 10^{-4}$ for $\rho=2.28$~g/cm$^3$ and $\rho=2.32$~g/cm$^3$, while $q_t$ plateaus off around $0.5$ at low shear rates for $\rho=2.44$~g/cm$^3$. This is a strong indication that, at low shear rates, the structural features of both LDL and HDL can be retained and that both liquid forms can be obtained under shear in the steady-state.

\begin{figure}
\begin{center}
\includegraphics*[width=12cm]{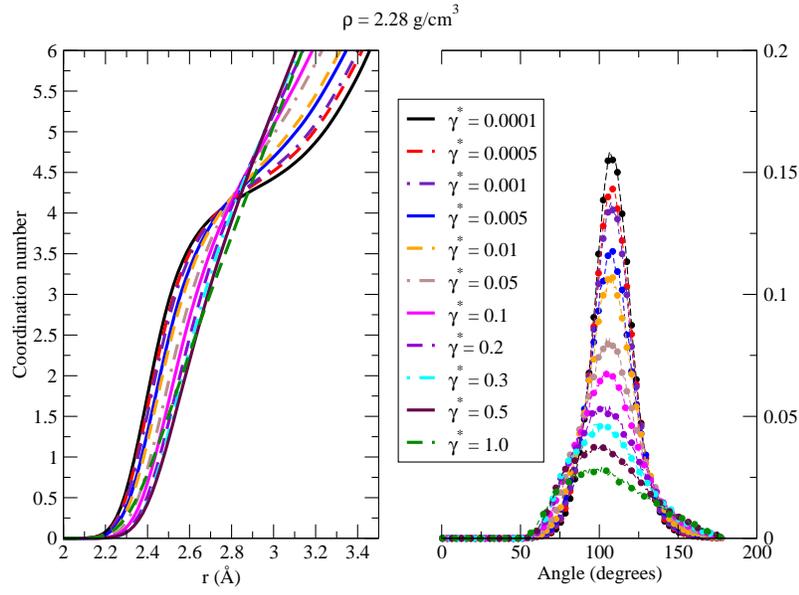}(a)
\includegraphics*[width=12cm]{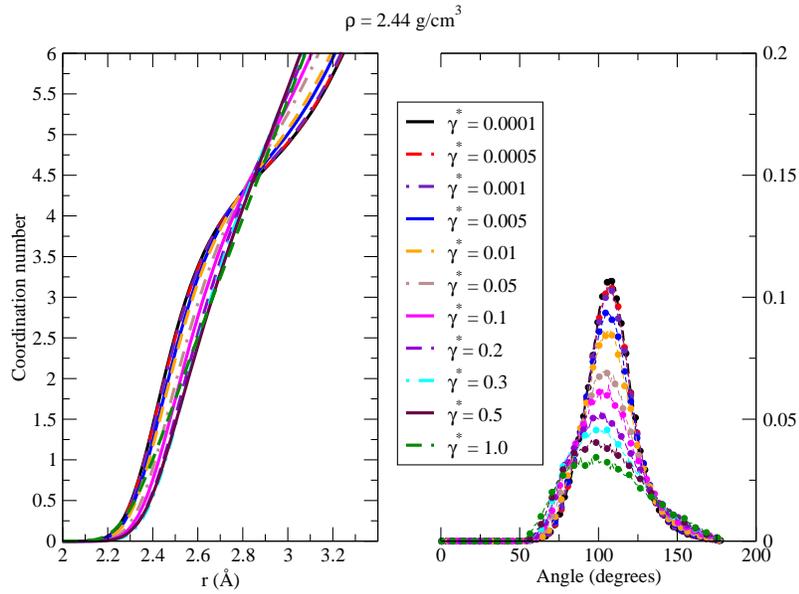}(b)
\end{center}
\caption{Structural features for (a) $\rho=2.28$~g/cm$^3$ and (b) $\rho=2.44$~g/cm$^3$ subjected to reduced shear rates ranging from $10^{-4}$ to $1$. For each plot, the left panel shows the coordination number $N_c$ as a function of the distance $r$, while the right panel shows the angle distribution for atoms with 4 first neighbors.}
\label{Fig4}
\end{figure}

To ascertain this further, we focus on the results obtained for $\rho=2.28$~g/cm$^3$ and $\rho=2.44$~g/cm$^3$ and examine how shear impacts the coordination number $N_c$ and the angle distributions. We start with the results for $\rho=2.28$~g/cm$^3$, shown in Fig.~\ref{Fig4}(a). We find that $N_c$ (left panel) exhibits two different behaviors as a function of the shear rate. For $\gamma^*<0.01$, the coordination number plot as a function of distance is similar to that observed at rest, with an inflection around $r=2.8$~\AA~associated with the two sharp peaks found for $g(r)$ for the LDL form. This is confirmed by the value obtained for the first  coordination shell ($Nc=4.3$ for $\gamma^*=0.0001$), which is consistent with that found at rest. As shown on Fig.~\ref{Fig4}(a), increasing the reduced shear rate beyond $0.01$ changes the shape of the coordination number plot, and, in turn, greatly reduces the amount of tetrahedral order in the fluid under shear as shown on the right panel of Fig.~\ref{Fig4}(a). Indeed, the maximum for the angle distribution decreases sharply with the shear rate, showing that fewer and fewer atoms have 4 first neighbors. Furthermore, the maximum for the distribution shifts towards the lower values at high shear rates and the distribution becomes tilted away from the distribution expected for a tetrahedral environment. The results therefore confirm that, provided that the reduced shear rate is below $0.01$, a steady-state of a liquid with the structural hallmarks of the LDL form can be stabilized using shear. Turning to the results for $\rho=2.44~g/cm^3$ in Fig.~\ref{Fig4}(b), we find that the coordination number and angle distribution for shear rates below $0.01$ are consistent with those found at rest. For instance, $N_c$ is found to be equal to $4.83$ for a shear rate of $1\times 10^{-4}$, in excellent agreement with the value obtained at rest. Similarly, as shown on the right panel of Fig.~\ref{Fig4}(b), the maximum for the angle distribution when $\gamma^*=1 \times 10^{-4}$ is $0.11$ in line with the value of $0.11$ found at rest. We also observe that shear rates beyond $0.01$ wipe away these features, as shown by the steady decrease in the maximum for the angle distribution with increasing shear. Overall, the results show that, for shear rates below $0.01$, we have succeeded in obtaining in the steady-state a liquid with the structural characteristics of the HDL form, and that both the steady-states of the two metastable forms of liquid Si have been stabilized under shear.

\begin{figure}
\begin{center}
\includegraphics*[width=9cm]{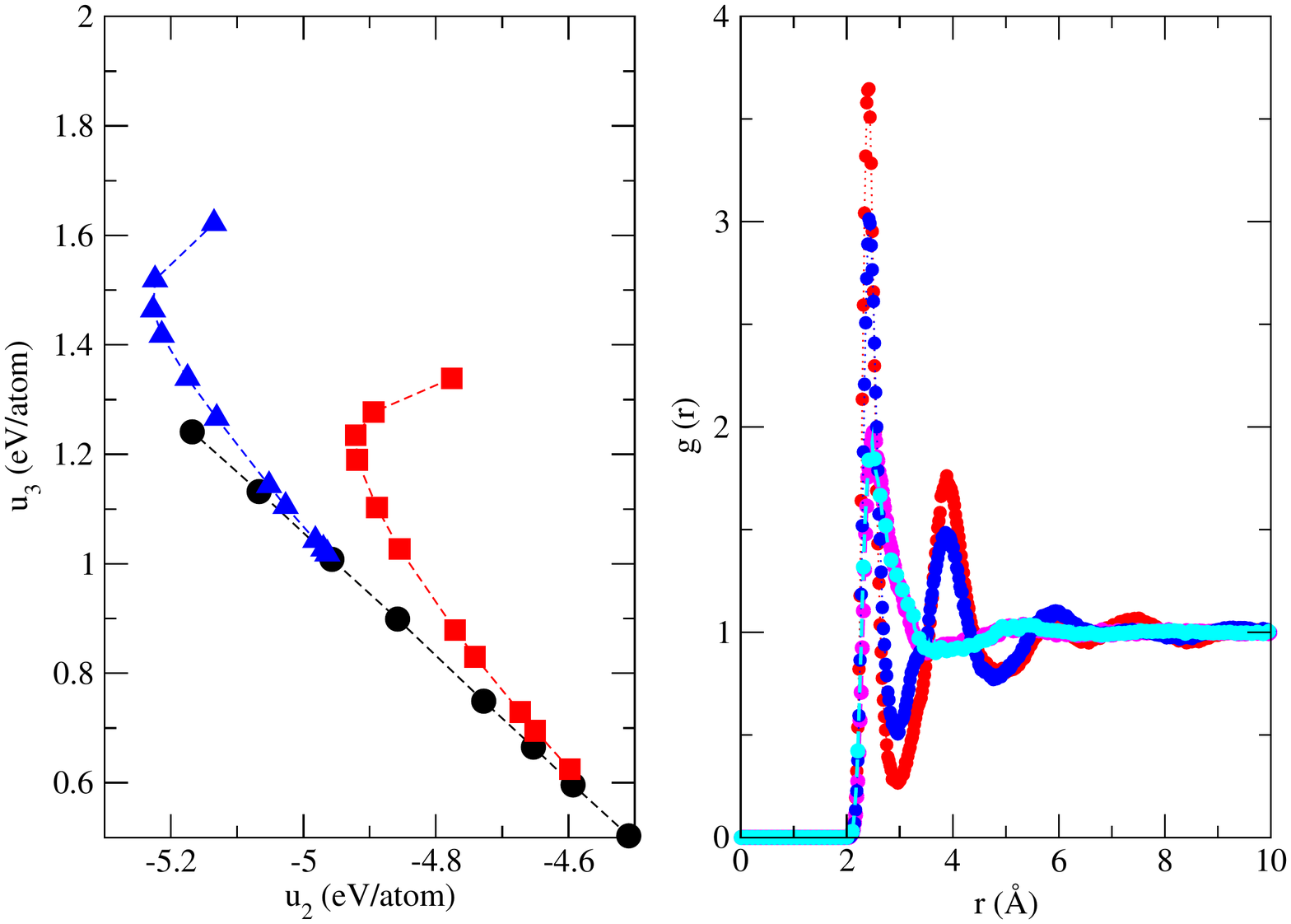}(a)
\includegraphics*[width=10.6cm]{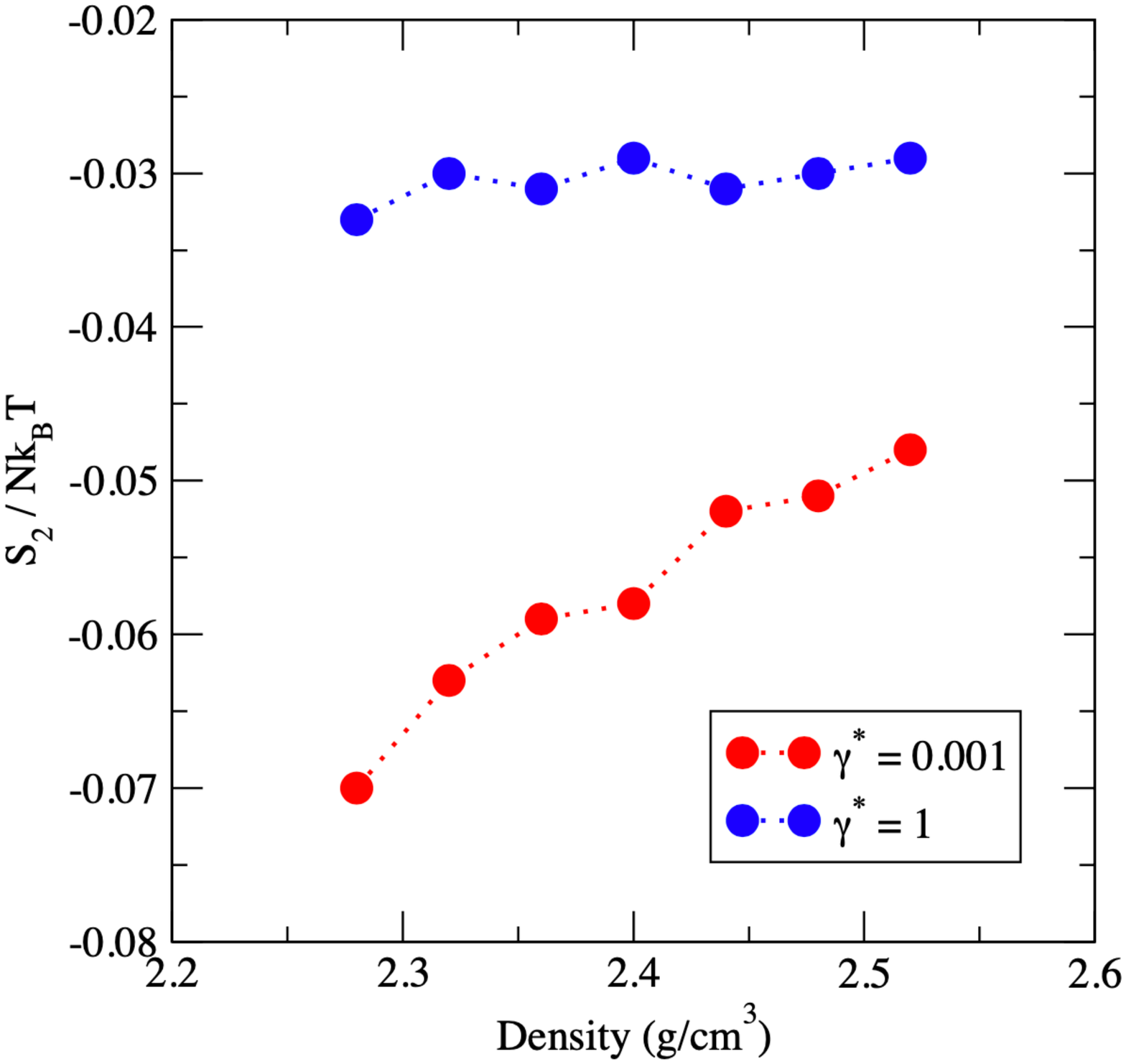}(b)
\end{center}
\caption{(a) (Left) $<u_3>$ vs. $<u_2>$ at rest for densities ranging from $2.28$~g/cm$^3$ (bottom right corner) to $2.44$~g/cm$^3$ (top left corner) is shown as circles. Also shown as red squares is $<u_3>$ vs. $<u_2>$ under shear for reduced shear rates from $1 \times 10^{-4}$ (bottom right corner) to $1$ (top left corner) at $\rho=2.28$~g/cm$^3$, while $<u_3>$ vs. $<u_2>$ under shear for reduced shear rates from $1 \times 10^{-4}$ (bottom right corner) to $1$ (top left corner) at $\rho=2.44$~g/cm$^3$ is shown as green triangles.(Right) $g(r)$ for $\rho=2.28$~g/cm$^3$ (red dashed line - bright red is used at low shear and dark red at high shear) and $\rho=2.44$~g/cm$^3$ (green dashed line - bright green is used at low shear and dark green at high shear). (b) $S_2$ against $\rho$ for a shear rate of $0.001$ (red circles) and a shear rate of $1$ (green circles)}
\label{Fig5}
\end{figure}

As discussed in Fig.~\ref{Fig2}(b), there is a linear relation between the two components of the potential energy that remains valid in both of the the LDL and HDL forms at rest. We compare in Fig.~\ref{Fig5}(a) the linear plot obtained at rest to the $<u_2>$ vs. $<u_3>$ plots obtained for different shear rates at $2.28$~g/cm$^3$ and $2.44$~g/cm$^3$. We observe that the results obtained for the lowest shear rates fall onto the $<u_2>$ vs. $<u_3>$ line obtained at rest, further establishing that shear rates below $0.01$ do not alter the nature of the LDL and HDL forms both from an energetic standpoint (Fig.~\ref{Fig5}(a)) and from a structural standpoint (Fig.~\ref{Fig4}). The energy plots for larger shear rates give some insight into the dramatic changes that take place at higher shear rates for both densities. As shear rate increases, the 2-body vs. 3-body relation departs more and more from the linear relation observed at rest, a trend that is confirmed by the structural changes experienced by the fluid at high shear rates as shown by the pair correlation functions obtained at high shear rates (Fig.~\ref{Fig5}(b)). While the $g(r)$ observed under low shear for $2.28$~g/cm$^3$ and $2.44$~g/cm$^3$ are in very good agreement with their counterparts at rest for the LDL and HDL forms (Fig.~\ref{Fig1}(a)), the $g(r)$ under high shear are very similar for both densities, with a single well-defined peak and little structural detail beyond. This confirms that applying too high a shear rate destabilizes the formation of tetrahedral order for all densities and prevents the system from exhibiting the two types of liquid forms obtained at rest. In other words, there is an upper limit to the applied shear (here identified to be $0.01$) that can be used to stabilize the LDL and HDL. 

Another way to assess the loss of organization, or, equivalently, of information at high shear can be made through the evaluation of the $S_2$ entropy. Fig.~\ref{Fig5}(b) shows a comparison of how $S_2$ varies with the liquid density for a shear rate of $0.001$ and a shear rate of $1$. We observe that, at $\gamma^*=0.001$ and for all densities, the liquid is more organized and has a lower $S_2$ than at high shear rate. This is in line with the less structured $g(r)$ shown in Fig.~\ref{Fig5}(a) for the larger shear rates. Furthermore, for a shear rate of $0.001$, $S_2$ increases with density. This can be attributed to the gradual loss of organization, that takes place as density increases as a result of the decrease in tetrahedral order. On the other hand,  $S_2$ is almost constant over the entire density range for $\gamma^*=1$. This stems from the very similar $g(r)$ obtained for all densities at high shear rates.

\begin{figure}
\begin{center}
\includegraphics*[width=12cm]{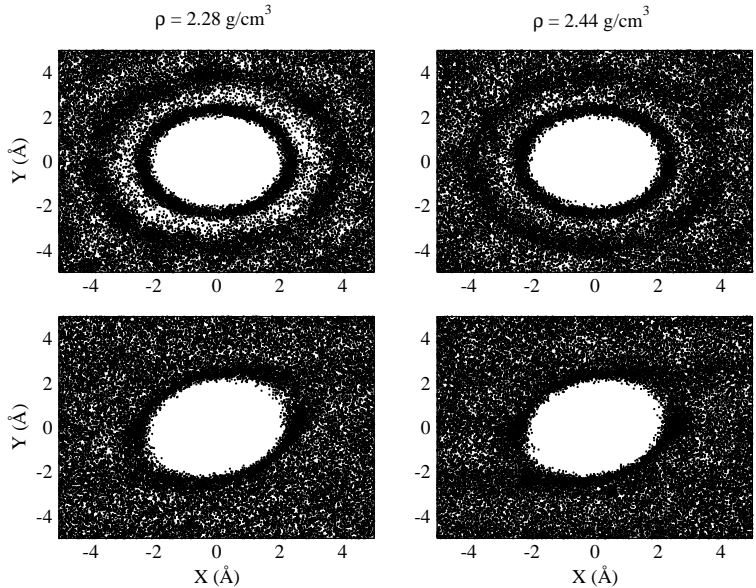}
\end{center}
\caption{Probability of finding a neighboring $Si$ atom in the $(x,y)$ plane within a slab of a width of $\sigma$ along the $z$ axis for $\rho=2.28$~g/cm$^3$ (left panel) and $\rho=2.44$~g/cm$^3$ (right panel). For each panel, the top graph corresponds to the lowest shear rate studied ($1 \times 10^{-4}$) and the bottom graph to the highest shear rate ($1$).}
\label{Fig6}
\end{figure}

To understand better the effect of shear on the liquid structure, we plot in Fig.~\ref{Fig6} a probability map of the presence of a neighboring atom in the $(x,y)$ plane within a slab of a width of $\sigma$ along the $z$ axis. At low shear rates, two dark circular regions appear clearly for both densities, corresponding to the first two peaks observed in the pair corrleation function. The contrast between the first two dark disks is much sharper at low density, as a result of the strong short-range tetrahedral order that takes place in the LDL form, than in the HDL form. This characterizes the in-plane structure of the liquids subjected to a low shear rate and confirms the LDL/HDL nature of the steady-state generated under these conditions. On the other hand, the plots are very different at high shear rate, with a single dark ellipse observed for all densities. This ellipse corresponds to the single peak exhibited by $g(r)$ at high shear. Furthermore, the ellipse clearly highlights the compression axis (diagonal that goes from the top left corner to the bottom right corner), which shows the increased contact, and this decreased distance between two Si atoms along that diagonal (this effect is due to the greater streaming velocity of, {\it e.g.}, an atom coming from the top left corner with respect to the central atom). Similarly, the opposite diagonal shows the elongation axis (bottom left corner to top right corner), with a greater distance between two neighboring atoms along that axis.  

\section{Conclusions}

In this work, we use NEMD methods to study the response of metastable liquids of Silicon, when subjected to shear. We show that, for sufficiently low shear rates, we achieve the formation, in the steady-state, of liquids that have similar structural, energetic and entropic signatures to the metastable liquids identified at rest as the low-density liquid (LDL) and high-density liquid (HDL) forms. In particular, we establish that, at $T=1090$~K, the LDL features are seen in Si under shear for densities below $2.4$~g/cm$^3$ and for reduced shear rates below than $0.01$, while the LDL features are recovered for densities greater than $2.4$~g/cm$^3$ and for reduced shear rates below than $0.01$. The competition between shear and tetrahedral ordering is also unraveled via the determination of the variations of the tetrahedral order parameter $q_t$ and of the two-body entropy $S_2$ as a function of the applied shear, leading to a cross-validation of the range of shear rates for which the two liquid polymorphs can be obtained. The results point to the efficiency and reliability of using shear as a means to stabilize metastable liquids under out-of-equilibrium conditions. Most notably, these nonequilibrium liquids often exhibit dramatically different properties. Indeed, the LDL of Si is semimetallic, with a pseudogap, while the HDL of Si is metallic, with no band gap. Being able to control the liquid properties via shear is an intriguing prospect, both for Si but also for the increasing range of atomic and molecular fluids, including water, that are polymorphic.

\begin{acknowledgments}
Partial funding for this research was provided by NSF through award CHE-1955403. This work used the Extreme Science and Engineering Discovery Environment (XSEDE)~\cite{xsede}, which is supported by National Science Foundation grant number ACI-1548562, and used the Open Science Grid through allocation TG-CHE200063.
\end{acknowledgments}

\vspace{1 cm}

{\bf Data availability}
\vspace{0.5 cm}

The data that support the findings of this study are available from the corresponding author upon reasonable request.

\bibliography{SiShear}

\end{document}